\documentclass{jpp}

\usepackage{graphicx}
\usepackage{natbib}
\usepackage[T1]{fontenc}
\usepackage{bm}
\usepackage[usenames, dvipsnames]{color}
\usepackage{graphicx}
\usepackage{amsmath}
\usepackage{amssymb}
\usepackage{amsfonts}
\usepackage{courier}
\usepackage{txfonts}
\usepackage[colorlinks=true,citecolor=blue,linkcolor=blue]{hyperref}

\ifCUPmtlplainloaded \else
  \checkfont{eurm10}
  \iffontfound
    \IfFileExists{upmath.sty}
      {\typeout{^^JFound AMS Euler Roman fonts on the system,
                   using the 'upmath' package.^^J}%
       \usepackage{upmath}}
      {\typeout{^^JFound AMS Euler Roman fonts on the system, but you
                   dont seem to have the}%
       \typeout{'upmath' package installed. JPP.cls can take advantage
                 of these fonts, if you use 'upmath' package.^^J}%
      }
  \else
  \fi
\fi

\ifCUPmtlplainloaded \else
  \checkfont{msam10}
  \iffontfound
    \IfFileExists{amssymb.sty}
      {\typeout{^^JFound AMS Symbol fonts on the system, using the
                'amssymb' package.^^J}%
       \usepackage{amssymb}%
       \let\le=\leqslant  
       \let\ge=\geqslant  
      }{}
  \fi
\fi

\ifCUPmtlplainloaded \else
  \IfFileExists{amsbsy.sty}
    {\typeout{^^JFound the 'amsbsy' package on the system, using it.^^J}%
     \usepackage{amsbsy}}
    {\providecommand\boldsymbol[1]{\mbox{\boldmath $##1$}}}
\fi

\newsavebox{\astrutbox}
\sbox{\astrutbox}{\rule[-5pt]{0pt}{20pt}}

\newcommand{\spa}{s_\|}
\newcommand{\spz}{s_{\| 0}}
\newcommand{\spe}{s_\perp}
\newcommand{\Eh}{\hat{E}}
\newcommand{\Zeff}{Z_{\rm eff}}
\newcommand{\Winf}{W_\infty}
\newcommand{\Ebar}{\bar{E}}
\newcommand{\Ebarm}{\bar{E}^{-1}}

\title[Non-monotonic runaway tail]{Radiation reaction induced non-monotonic features in runaway electron distributions}

\author[E. Hirvijoki, I. Pusztai, J. Decker, O. Embr\'{e}us, A. Stahl and T. F\"{u}l\"{o}p]%
{E.\ns H\ls I\ls R\ls V\ls I\ls J\ls O\ls K\ls I$^1$%
  \thanks{Email address for correspondence: eero.hirvijoki@chalmers.se},\ns
I.\ns P\ls U\ls S\ls Z\ls T\ls A\ls I$^1$,\ns
J.\ns D\ls E\ls C\ls K\ls E\ls R$^2$,
\ns
O.\ns E\ls M\ls B\ls R\ls \'{E}\ls U\ls S$^1$,
\ns
A.\ns S\ls T\ls A\ls H\ls L$^1$ \ns
\and T.\ns F\ls \"{U}\ls L\ls \"{O}\ls P$^1$}

\affiliation{$^1$Department of Applied Physics, Chalmers University of 
Technology, 41296 Gothenburg, Sweden\\[\affilskip]
$^2$Ecole Polytechnique F\'{e}d\'{e}rale de Lausanne (EPFL), Centre de 
Recherches en Physique des Plasmas (CRPP), CH-1015 Lausanne, Switzerland}

\pubyear{?}
\volume{?}
\pagerange{?}
\date{?; revised ?; accepted ?. - To be entered by editorial office}
\begin{document}

\maketitle

\begin{abstract}
Runaway electrons, which are generated in a plasma where the induced
electric field exceeds a certain critical value, can reach very high
energies in the $\rm MeV$ range. For such energetic electrons,
radiative losses will contribute significantly to the momentum space
dynamics. Under certain conditions, due to radiative momentum losses,
a non-monotonic feature -- a ``bump'' -- can form in the runaway
electron tail, creating a potential for bump-on-tail-type
instabilities to arise. Here we study the conditions for the existence
of the bump. We derive an analytical threshold condition for bump
appearance and give an approximate expression for the minimum energy
at which the bump can appear. Numerical calculations are performed to
support the analytical derivations.
\end{abstract}


\section{Introduction}
\label{sec:intro}
In a plasma, the drag force from Coulomb collisions acting on fast
electrons decreases with the electron velocity. Thus, if the electric
field $E$ exceeds a threshold value $E_{c}$, electrons with sufficient
velocity will be indefinitely accelerated and are called \emph{runaway
  electrons}. The critical field $E_{c}$ is defined as
\begin{equation}
E_{c}=\frac{n_ee^{3}\ln\Lambda}{4\pi\varepsilon_{0}^2m_ec^{2}}, \label{eq:critical_field}
\end{equation}
where $n_e$ is the electron density, $m_e$ is the electron rest mass,
$c$ is the speed of light, $e$ is the elementary charge, and
$\ln\Lambda$ is the Coulomb logarithm.

Runaway electrons are generated in the presence of an induced electric
field parallel to the magnetic field. In a tokamak, the condition
$E>E_c$ can be met during the plasma start-up, during the flat-top
phase of Ohmic plasmas if the density is sufficiently low, or in
plasma disruptions. Especially during disruption events, a beam of
runaways carrying a current of several MA and an energy of several $\rm MJ$,
may form. Such a runaway beam would pose a serious threat to the
integrity of the first wall in reactor-size fusion devices. Any
mechanism that could possibly limit the formation of a considerable
runaway beam would be of importance. 

While the role of radiative momentum losses due to synchrotron
emission has been studied previously in \citep{andersson:pop2001}, the
possibility of a non-monotonic feature in the energy distribution of
runaways -- which we will henceforth refer to as a ``bump'' -- was not
considered. The formulation of the problem in
\citep{andersson:pop2001} does not ensure particle conservation in the
presence of radiation reaction and neglects certain terms needed to
describe the bump \citep{hazeltine2004,stahl:prl}. The possibility of
bump formation, however, immediately raises the question of whether
the non-monotonic behavior of the distribution could lead to kinetic
instabilities, causing a redistribution of the runaway particles,
favorable for mitigating the potential threat to the machine. A
thorough investigation of conditions favoring bump formation is thus
needed.

In the present paper, we use analytical calculations to investigate
the runaway electron distribution under the combined influence of
Coulomb collisions, electric field acceleration, and radiative
momentum losses. We show the existence of a bump and derive both a
threshold condition for the appearance of the bump and an approximate
expression for its location in parallel momentum space. The accuracy
of the analytical estimates are then tested against numerical
simulations carried out using CODE \citep{Landreman2014847,stahl:prl}.

The paper is organized as follows. In Sec.~\ref{sec:kinetic_equation},
we start by describing the particle phase-space kinetic equation. We also discuss the
transformation of the kinetic equation into the guiding-center
phase-space and give the corresponding expression in the case of a
uniform plasma. Analytical calculation of the condition for bump-on-tail appearance,
based on the guiding-center dynamics, are presented in Sec.~\ref{sec:characteristics}. A comparison of the derived conditions to numerical results is presented in
Sec.~\ref{sec:numres}, before we conclude in
Sec.~\ref{sec:conclusions}.

\section{Kinetic equation}
\label{sec:kinetic_equation}
The kinetic equation describing the dynamics of charged particles in a
plasma is
\begin{align}
\frac{\partial f_a}{\partial
  t}+\frac{\partial}{\partial\mathbf{x}}\cdot\left(\dot{\mathbf{x}}f_a\right)+\frac{\partial}{\partial\mathbf{p}}\cdot\left(\dot{\mathbf{p}}f_a\right)=C[f_a,f_b],
\end{align}
where $C[f_a,f_b]$ is the collision operator for collisions between
particle species $a$ and $b$, $z^{\alpha}=(\mathbf{x},\mathbf{p})$
are the phase-space coordinates, and
$\dot{z}^{\alpha}=(\dot{\mathbf{x}},\dot{\mathbf{p}})$ the equations
of motion. In the Fokker-Planck limit, the Coulomb collision operator
is given by
 \begin{align}
C[f_a,f_b]=-\frac{\partial}{\partial\mathbf{p}}\cdot\left(\mathbf{K}_{ab}[f_b]f_a-\mathbb{D}_{ab}[f_b]\cdot\frac{\partial f_a}{\partial\mathbf{p}}\right), \label{eq:particle_phase_space_coll_op}
\end{align}
where $\mathbf{K}_{ab}[f_b]$ is the friction vector and
$\mathbb{D}_{ab}[f_b]$ the diffusion tensor (see
Appendix~\ref{app:relativistic_collision_operator} for details). In
this paper, we do not consider contributions from large-angle
collisions.

The equations of motion for a particle with charge $q$ and mass $m$
combine the Hamiltonian motion from the electric and magnetic fields
$\mathbf{E}$ and $\mathbf{B}$, and a force $\mathbf{F}$
that accounts for non-Hamiltonian dynamics
\begin{align}
\dot{\mathbf{x}}=&\mathbf{v},\\
\dot{\mathbf{p}}=&q\mathbf{E}+q\,\mathbf{v}\times\mathbf{B}+\mathbf{F}.
\end{align}
Here $\mathbf{p}=\gamma m\mathbf{v}$ is the particle momentum, and
\mbox{$\gamma=1/\sqrt{1-v^{2}/c^{2}}=\sqrt{1+p^{2}/(mc)^{2}}$} is the
relativistic factor. In the case considered here, the non-Hamiltonian
force is the radiation reaction (RR) force which was first described
by~\citet{lorentz2} in the case of a classical non-relativistic point
charge, and was later generalized to relativistic energies
by~\citet{abraham} and~\citet{dirac:1938}. As such, the
Lorentz-Abraham-Dirac (LAD) force is~\citep{pauli1958theory}
\begin{align}\label{eq:reaction_force}
\mathbf{F}_{\textrm{LAD}}=\frac{q^{2}\gamma^{2}}{6\pi\varepsilon_{0}c^{3}}
\left[\ddot{\mathbf{v}}+\frac{3\gamma^{2}}{c^{2}}\left(\mathbf{v}\cdot
\dot{\mathbf{v}}\right)\dot{\mathbf{v}}+\frac{\gamma^{2}}{c^{2}}
\left(\mathbf{v}\cdot\ddot{\mathbf{v}}+\frac{3\gamma^{2}}{c^{2}}
\left(\mathbf{v}\cdot\dot{\mathbf{v}}\right)^{2}\right)\mathbf{v}\right].
\end{align}

\noindent The LAD-force does however contain third order time
derivatives of the particle position, which allows for the existence
of pathological solutions. For instance, the particle velocity may
grow exponentially in the absence of external forces ($\mathbf{E}=0$,
$\mathbf{B}=0$), see e.g. \citep{rohrlich2007classical}. These
issues have generated discussion regarding which expression to use for
the RR-force.  \citet{landau_lifshitz_fields} suggested a perturbative
approach in which the velocity derivatives in
Eq.~(\ref{eq:reaction_force}) are expressed in terms of the external
force only (here the Lorentz force). \citet{Ford1993182} argue that
this approach is in fact the correct one. In the paper
by~\citet{Spohn:EPL:2000}, it is shown that the non-physical solutions
can be avoided if the LAD-force is limited on a so-called {\em
  critical surface}, and that the resulting expressions will be
equivalent to those of the perturbative approach. We have thus chosen
to adopt the perturbative approach. Furthermore, we neglect the
electric field in the expressions for $\dot{\mathbf{v}}$ and
$\ddot{\mathbf{v}}$ in the RR-force.  This is justified since the
motion of the particle is dominated by the magnetic field in the
strongly magnetized plasmas considered here.  An excellent discussion
about the RR-force can be found in a recent review paper
by~\citet{RevModPhys.84.1177}.

\subsection{Guiding-center transformation}
Because of the $\mathbf{v}\times\mathbf{B}$--term, the particle
phase-space kinetic equation in a magnetized plasma includes the rapid
gyromotion time-scale which is often not interesting and is expensive
to resolve computationally. It can, however, be eliminated using
guiding-center Lie-transform perturbation methods. The transformation
of the Hamiltonian equations of motion is one of the classical results
in modern plasma physics
\citep[see][]{littlejohn:jpp:4732464,RevModPhys.81.693}, and the
Fokker-Planck collision operator has been considered
in~\citep{brizard:4429,decker:112513,hirvijoki:092505}. The final step
necessary to formulate our problem, the transformation of the RR-force,
was given recently in~\citep{hirvijoki:2014}.

After the transformation, the guiding-center kinetic equation for a
gyro-angle averaged distribution function $\left\langle
F_a\right\rangle$, including Hamiltonian motion in electromagnetic
fields, the RR-force, and Coulomb collisions in the Fokker-Planck limit,
is given by
\begin{align}
\label{eq:gc_kinetic}
\frac{\partial\left\langle F_a\right\rangle }{\partial t}+\frac{1}{\mathcal{J}}\frac{\partial}{\partial Z^{\alpha}}\left[\mathcal{J}\left(\frac{}{}\dot{Z}^{\alpha}+\left\langle\mathcal{F}_{gc\textrm{RR}}^{\alpha}\right\rangle\right)\left\langle F_a\right\rangle \right]=C_{gc\textrm{FP}}[\left\langle F_a\right\rangle],
\end{align}
where $Z^{\alpha}$ form the 5D guiding-center phase-space,
$\mathcal{J}$ is the guiding-center phase-space Jacobian,
$\dot{Z}^{\alpha}$ are the Hamiltonian guiding-center equations of
motion, and
$\left\langle\mathcal{F}_{gc\textrm{RR}}^{\alpha}\right\rangle$ is
the contribution from the RR-force to the guiding-center
motion. Similarly to the particle phase-space operator in Eq.~\eqref{eq:particle_phase_space_coll_op}, we can write
the guiding-center Fokker-Planck collision operator in phase-space
divergence form
\begin{align}
\label{eq:full_gc_collision_operator}
C_{gc\textrm{FP}}[\left\langle F_a\right\rangle]=-\frac{1}{\mathcal{J}}\frac{\partial}{\partial Z^{\alpha}}\left[\mathcal{J}\left(\left\langle\mathcal{K}_{ab,gc}^{\alpha}\right\rangle\left\langle F_a\right\rangle-\left\langle\mathcal{D}_{ab,gc}^{\alpha\beta}\right\rangle\frac{\partial\left\langle F_a\right\rangle}{\partial Z^{\beta}}\right)\right],
\end{align}
where $\left\langle\mathcal{K}_{ab,gc}^{\alpha}\right\rangle$ and
$\left\langle\mathcal{D}_{ab,gc}^{\alpha\beta}\right\rangle$ are the
guiding-center Coulomb friction and diffusion coefficients.

\subsection{Equations of motion}
We solve Eq.~(\ref{eq:gc_kinetic}) in a uniform plasma, using 2D
guiding-center velocity space coordinates $Z^{\alpha}=(p,\xi)$, where $p$
is the absolute value of the guiding-center momentum,
$\xi=p_{\parallel}/p$ is the pitch-angle-cosine ($p_{\parallel}$ is
the guiding-center momentum parallel to the magnetic field) and the 
guiding-center Jacobian is given by $\mathcal{J}=p^2$. In this
case, the guiding-center equations of motion take the simple forms
\begin{align}
\dot{p}=&qE_{\parallel}\xi,\\
\dot{\xi}=&qE_{\parallel}(1-\xi^2)/p,
\end{align}
where $E_{\parallel}$ is the electric field parallel to the magnetic
field. The components of the guiding-center RR-force in the limit 
corresponding to pure synchrotron emission are \citep{hirvijoki:2014}
\begin{align}
\left\langle\mathcal{F}_{gc\textrm{RR}}^p\right\rangle=&-\frac{\gamma p(1-\xi^2)}{\tau_r},\\
\left\langle\mathcal{F}_{gc\textrm{RR}}^{\xi}\right\rangle=&\;\frac{\xi(1-\xi^2)}{\gamma\tau_r},
\end{align}
where the radiation reaction time-scale is defined by
\begin{align}
\tau_{r}=\frac{6\pi\varepsilon_{0}(mc)^{3}}{q^{4}B^{2}}=\frac{3c}{2r\gamma^2\Omega^2},
\end{align}
with $r=q^2/(4\pi\varepsilon_0mc^2)$ the classical electron radius and
\mbox{$\Omega=qB/(\gamma m)$} the gyro-frequency.

\subsection{Collision operator}
The particle phase-space friction and diffusion coefficients,
$\mathbf{K}_{ab}[f_b]$ and $\mathbb{D}_{ab}[f_b]$, are expressed in
terms of the five relativistic Braams-Karney potential functions,
which are weighted integrals of the background distribution functions
$f_b$; \citep[see][]{braams_karney:1989} and
Appendix~\ref{app:relativistic_collision_operator} for details. If
the particle species $a$ and $b$ coincide, the self-collisions result
in a nonlinear collision operator. In the present study, the particle
phase-space collision operator is transformed into the guiding-center
phase-space and linearized around a Maxwellian. The integral terms of
the linearized collision operator are neglected and only the test
particle contribution is considered. This choice, with some further
simplifications, allows analytical solution of
Eq.~(\ref{eq:gc_kinetic}), which will be discussed in
Sec.~\ref{sec:characteristics}.

The guiding-center friction and diffusion coefficients
$\left\langle\mathcal{K}_{ab,gc}^{\alpha}\right\rangle$ and
$\left\langle\mathcal{D}_{ab,gc}^{\alpha\beta}\right\rangle$ that
appear in the guiding-center Fokker-Planck operator in
Eq.~(\ref{eq:full_gc_collision_operator}) are gyro-averaged
projections of their guiding-center push-forwarded particle phase-space
counterparts. For a detailed definition of the guiding-center friction
and diffusion coefficients, we refer
to~\citep{brizard:4429,decker:112513,hirvijoki:092505}. The general
expressions are non-trivial but, in the limit of a uniform plasma, the
test-particle operator assuming isotropic background particle
distributions becomes diagonal with reasonably simple non-zero
components
\begin{align}
\left\langle\mathcal{K}_{ab,gc}^p\right\rangle&\equiv-\nu_{l,ab}\, p,\\
\left\langle\mathcal{D}_{ab,gc}^{pp}\right\rangle&\equiv\;D_{l,ab},\\
\left\langle\mathcal{D}_{ab,gc}^{\xi\xi}\right\rangle&\equiv\;(1-\xi^2)\frac{D_{t,ab}}{p^2}.
\end{align}
The coefficients $\nu_{l,ab}$, $D_{l,ab}$, and $D_{t,ab}$, where the sub-indices $l$ and
$t$ stand for ``longitudinal'' and ``transverse'' with respect to the
guiding-center momentum vector, are expressed
in terms of the five Braams-Karney potentials $\Psi_n(u)$ with
$u=p/m_a$ and $\Gamma_{ab}=q_a^2q_b^2\ln\Lambda/(4\pi\varepsilon_0^2)$
according to
\begin{align}
\nu_{l,ab}=&\;4\pi\frac{m_a}{m_b}\Gamma_{ab}\frac{\gamma}{p}\left(\frac{\partial\Psi_1}{\partial u}-\frac{2}{c^2}\frac{\partial\Psi_2}{\partial u}\right),\\
D_{l,ab}=&-4\pi\Gamma_{ab}\gamma\left(\Psi_0-\frac{2\gamma^2}{u}\frac{\partial\Psi_3}{\partial u}+\frac{8\gamma^2}{uc^2}\frac{\partial\Psi_4}{\partial u}-\frac{8}{c^4}\Psi_4\right),\\
D_{t,ab}=&-4\pi\Gamma_{ab}\gamma\left(\frac{1}{u}\frac{\partial\Psi_3}{\partial u}+\frac{1}{c^2}\Psi_3-\frac{4}{uc^2}\frac{\partial\Psi_4}{\partial u}+\frac{4}{c^4}\Psi_4\right).
\end{align}
Our guiding-center Fokker-Planck operator thus becomes
\begin{equation}
C_{gc\textrm{FP}}[\left\langle F_a\right\rangle]=\frac{1}{p^2}\frac{\partial}{\partial p}\left[p^2\left(\nu_{l,ab}\,p\left\langle F_a\right\rangle+D_{l,ab}\frac{\partial\left\langle F_a\right\rangle}{\partial p}\right)\right] +\frac{D_{t,ab}}{p^2}\frac{\partial}{\partial\xi}\left[\left(1-\xi^2\right)\frac{\partial\left\langle F_a\right\rangle}{\partial\xi}\right],
\end{equation}
where the first term with momentum derivatives is responsible for the
slowing down of fast particles and momentum diffusion, while the
second term describes scattering in pitch-angle. 

\subsection{Final expression}
For the rest of the paper, to streamline notation, we shall suppress
the brackets that denote the gyro-averaging and the sub-index from the
expression for the parallel electric field. Also, we will drop the
particle species indices as we sum over all the background species in
the collision operator. Thus we will have $\nu_l=\sum_{b}\nu_{l,ab}$,
$D_l=\sum_bD_{l,ab}$, and $D_t=\sum_bD_{t,ab}$, and our kinetic
equation in the continuity form becomes
\begin{align}
\frac{\partial F}{\partial t}+&\frac{1}{p^2}\frac{\partial}{\partial p}\left[p^2\left(qE\xi-\frac{\gamma p(1-\xi^2)}{\tau_r}-\nu_lp\right)F-p^2D_l\frac{\partial F}{\partial p}\right]\nonumber\\+&\frac{\partial}{\partial\xi}\left[(1-\xi^2)\left(\frac{qE}{p}F+\frac{\xi}{\gamma\tau_r}F-\frac{D_{t}}{p^2}\frac{\partial F}{\partial\xi}\right)\right]=0.
\label{eq:final_kinetic_equation}
\end{align}

In the following, we analyze this equation in detail. We describe
the formation of a bump-on-tail in the electron distribution function
both analytically and numerically. We also study the threshold
conditions for the bump formation and the minimum energy of the bump
location.

\section{Characteristics of a bump-on-tail feature}
\label{sec:characteristics}

The RR-force in a straight magnetic field system increases with
  the square of the perpendicular momentum, $\spe^2=s^2(1-\xi^2)$. As
  a consequence, the extent of the distribution function will, qualitatively, 
  be limited in $\spe$ to a region where the
  parallel component of the total force acting on an electron is
  positive. Electrons with higher perpendicular momenta are
  decelerated since the radiation reaction force overcomes the
  acceleration due to the parallel electric field. Compared to the
  case without RR-force, where the distribution function is
  continuously expanding in $\spe$ for increasing values of the
  parallel momentum $\spa=\xi s$, the limited extent of the
  distribution in $\spe$ when the RR-force is included 
  leads to qualitatively different dynamics.
  
The width of the distribution in $\spe$ is approximately
  constant, which means that pitch angle scattering is increasingly
  more effective at higher $\spa$ in moving the electrons to the
  region of phase space where they are decelerated. A consequence of
  this is that a true steady state solution of the kinetic equation
  exists and the distribution function decays exponentially in the far
  tail, something which was also observed in previous works, such as
  \citep{andersson:pop2001}.  Another new feature is the possibility
  of non-monotonic behavior in the tail of the steady state distribution
  function.  Note that this feature cannot be correctly described if
  the RR-force is not implemented in the phase-space divergence form
  that conserves the phase-space density.  Therefore it is
  overlooked by some earlier studies. To understand the properties of 
  the bump, and its formation, we will start the following analysis by
  assuming that a bump exists in the runaway tail, and make assumptions
  regarding its properties. These assumptions will be justified \emph{a
    posteriori} when our results are compared to numerical results in
  Sec.~\ref{sec:numres}.
 
Considering a possible bump-on-tail scenario, we study
Eq.~(\ref{eq:final_kinetic_equation}) in a region where the electrons
have high velocities compared to the electron and ion thermal speeds
\mbox{$v\gg v_{th,e},v_{th,i}$}.  Then the slowing-down force is
dominated by electron-electron collisions and it overshadows momentum
diffusion. For pitch-angle-scattering, collisions with both ions and
the electron bulk are important.

In the limit where the bulk populations are non-relativistic
Maxwellians, we have for the friction coefficient at high speeds
\begin{align}
\nu_l\approx \frac{n_ee^4\ln\Lambda}{4\pi\varepsilon_0^2m_ev^2}\frac{1}{p}\equiv\frac{eE_c}{\beta^2p},
\end{align}
and similarly for the transverse diffusion coefficient
\begin{align}
D_t\approx\frac{1+Z_{\textrm{eff}}}{2}\frac{n_ee^4\ln\Lambda}{4\pi\varepsilon_0^2}\frac{1}{v}\equiv\frac{1+Z_{\textrm{eff}}}{2}\frac{eE_cm_ec}{\beta},
\end{align}
where $E_c$ is the critical electric field (Eq.~\ref{eq:critical_field}), $Z_{\textrm{eff}}$ is the
effective ion charge, and $\beta=v/c$. These estimates coincide with
the expressions in~\citep{andersson:pop2001}. We define the normalized
momentum $s=p/(m_e c)$, time $\tau=eE_ct/(m_e c)$, radiation reaction
time-scale $\sigma^{-1}=eE_c\tau_r/(m_ec)$, and electric field
$\hat{E}=-E/E_c$, and transform the kinetic equation into a
dimensionless form for further analysis
\begin{align}
\frac{\partial F}{\partial\tau}+&\frac{1}{s^2}\frac{\partial}{\partial s}\left[s^2\left(\hat{E}\xi-\sigma\gamma s(1-\xi^2)-\frac{\gamma^2}{s^2}\right)F\right]\nonumber\\+&\frac{\partial}{\partial\xi}\left[(1-\xi^2)\left(\hat{E}\frac{F}{s}+\frac{\sigma\xi}{\gamma}F-\frac{\gamma}{s}\frac{1+Z_{\textrm{eff}}}{2s^2}\frac{\partial F}{\partial\xi}\right)\right]=0.
\end{align}

As the electric field affects only the parallel acceleration we expect
the system to be strongly biased about $\xi=1$. The phase-space volume
element in $(s,\xi)$--coordinates ($\mathcal{J}\sim s^2$), however,
scales nonlinearly close to $\xi\approx 1$. A better choice for
further studies close to the $\xi=1$ region is to use coordinates
$(s_{\parallel},s_{\perp})$ which have a Jacobian $\mathcal{J}\sim
s_{\perp}$ that stays constant with respect to $s_{\parallel}$. The
new coordinates relate to $(s,\xi)$ according to
\begin{align}
s_{\parallel}=&\;s\xi,\\
s_{\perp}=&\;s\sqrt{1-\xi^2}.
\end{align}
and our kinetic equation expressed with $(s_{\parallel},s_{\perp})$ becomes
\begin{align}
\frac{\partial F}{\partial\tau}+&\hat{E}\frac{\partial F}{\partial
  s_{\parallel}}-\frac{2F}{s}-\frac{\gamma^2}{s^2}\left(\frac{s_{\parallel}}{s}
\frac{\partial F}{\partial
  s_{\parallel}}+\frac{s_{\perp}}{s}\frac{\partial F}{\partial
  s_{\perp}}\right)
\nonumber\\ -&\frac{\gamma(1+Z_{\textrm{eff}})}{2s}\left[\frac{1}{s_{\perp}}
  \frac{\partial}{\partial s_{\perp}}\left(s_{\perp}\frac{\partial
    F}{\partial
    s_{\perp}}\right)+\frac{s_{\perp}^2}{s^2}\left(\frac{\partial^2F}{\partial
    s_{\parallel}^2}-\frac{\partial^2F}{\partial
    s_{\perp}^2}\right)\right.
- 2\left.\frac{s_{\parallel}s_{\perp}}{s^2}
  \frac{\partial^2F}{\partial s_{\parallel}\partial
    s_{\perp}}-2\frac{s_{\parallel}}{s^2}\left(\frac{\partial
    F}{\partial
    s_{\parallel}}+\frac{s_{\perp}}{s_{\parallel}}\frac{\partial
    F}{\partial
    s_{\perp}}\right)\right]\nonumber\\-&\frac{\sigma}{\gamma}\left[(2+4s_{\perp}^2)
  F+s_{\perp}(1+s_{\perp}^2)\frac{\partial F}{\partial
    s_{\perp}}+s_{\parallel}s_{\perp}^2\frac{\partial F}{\partial
    s_{\parallel}}\right]=0.
\label{eq:kineticinpape}
\end{align}
Instead of attempting to solve Eq.~(\ref{eq:kineticinpape}), in the
following we will concentrate on the dynamics at $\spe=0$, which will
be sufficient to prove the existence of a bump and to estimate its 
location in the electron tail.

We assume the distribution to be a smooth function of $s_{\perp}$,
which allows us to create a power series expansion around
$s_{\perp}=0$:
\begin{equation}
F(s_{\parallel},s_{\perp})=\sum_{n=0}^{\infty}\frac{s_{\perp}^{2n}}{(2n)!}\left[\frac{\partial^{(2n)} F}{\partial s_{\perp}^{(2n)}}\right]_{(s_{\parallel},0)}+s_{\perp}\sum_{n=0}^{\infty}\frac{s_{\perp}^{2n}}{(2n+1)!}\left[\frac{\partial^{(2n+1)} F}{\partial s_{\perp}^{(2n+1)}}\right]_{(s_{\parallel},0)}.
\end{equation}
Because the electric field is acting only in the parallel direction
$F$ is ''even'' in $s_{\perp}$, i.e., we can formally state that
$F(s_{\parallel},s_{\perp})=F(s_{\parallel},-s_{\perp})$ although our
phase-space does not extend to $s_{\perp}<0$. Thus, all the odd
$s_{\perp}$-derivatives at $s_{\perp}=0$ must vanish, and we find
\begin{align}
F(s_{\parallel},s_{\perp})\equiv\sum_{n=0}^{\infty}\frac{s_{\perp}^{2n}}{(2n)!}\left[\frac{\partial^{(2n)} F}{\partial s_{\perp}^{(2n)}}\right]_{(s_{\parallel},0)}.
\end{align}
With the help of the expansion, we may accurately calculate the limit
\begin{align}
\lim_{s_{\perp}\rightarrow 0}\frac{1}{s_{\perp}}\frac{\partial}{\partial s_{\perp}}\left(s_{\perp}\frac{\partial F}{\partial s_{\perp}}\right)=2\left[\frac{\partial^{2} F}{\partial s_{\perp}^{2}}\right]_{(s_{\parallel},0)},
\end{align}
and write the $s_{\perp}=0$ limit of the kinetic equation as 
\begin{align}
\label{eq:accurate_limit}
\left[\frac{\partial F}{\partial\tau}\right]_{(s_{\parallel},0)}+&\left(\hat{E}-\frac{1+s_{\parallel}^2}{s_{\parallel}^2}+\frac{(1+Z_{\textrm{eff}})\sqrt{1+s_{\parallel}^2}}{s_{\parallel}^2}\right)\left[\frac{\partial F}{\partial s_{\parallel}}\right]_{(s_{\parallel},0)} \\ -&\frac{(1+Z_{\textrm{eff}})\sqrt{1+s_{\parallel}^2}}{s_{\parallel}}\left[\frac{\partial^{2} F}{\partial s_{\perp}^{2}}\right]_{(s_{\parallel},0)}-2\left(\frac{\sigma}{\sqrt{1+s_{\parallel}^2}}+\frac{1}{s_{\parallel}}\right) \left[F\right]_{(s_{\parallel},0)}=0.
\nonumber
\end{align}
Assuming that a steady state solution exists, the possible extrema
are characterized by the condition
\begin{align}
\left[\frac{\partial F}{\partial s_{\parallel}}\right]_{(s_{\parallel},0)}=\;0.
\end{align}
We thus find an algebraic equation that defines the locations of these extrema
\begin{align}
\label{eq:extremaequation}
2\left(\frac{\sigma}{\sqrt{1+s_{\parallel}^2}}+\frac{1}{s_{\parallel}}\right)+\frac{(1+Z_{\textrm{eff}})\sqrt{1+s_{\parallel}^2}}{s_{\parallel}}\left[\frac{1}{F}\frac{\partial^{2} F}{\partial s_{\perp}^{2}}\right]_{(s_{\parallel},0)}=0.
\end{align}

\subsection{Threshold condition for the appearance of the bump}
Considering a steady-state solution to Eq.~(\ref{eq:accurate_limit}),
in a situation where the bump is on the verge of appearing, a single
inflection point exists in the distribution function instead of local maxima or minima. In this
section we derive a threshold condition describing the appearance of
an inflection point, by requiring the first and second
$\spa$-derivatives of the distribution to vanish simultaneously. 

Before we start the analysis, we note that the steady state
distribution function represented by Eq.~(12) of~\citep{andersson:pop2001} is separable in $\spa$ and $\spe$, and
it is of the form $\propto\exp[-\Winf^2 \spe^2/2]$, where
$\Winf^2=2\sigma/(\Eh-1)$. To find this result they
neglect $f/s_\|$ corrections compared to $\partial f/\partial s_\|$
terms in the kinetic equation, which is appropriate in the very far
tail [$s_\|$ corresponds to $p_\|$ in the
notation of~\citep{andersson:pop2001}]. Therefore,  the quantity
\begin{align}
W^2(\spa)\equiv-\left[\frac{1}{F}\frac{\partial^{2} F}{\partial s_{\perp}^{2}}\right]_{(s_{\parallel},0)}
\end{align} 
should approach $\Winf^2$ in the $\spa\rightarrow \infty$ limit. Thus
it is useful to define $\kappa(\spa)$, so that
$W^2(\spa)=\kappa(\spa)\Winf^2$ and $\kappa\rightarrow 1$ as
$\spa\rightarrow\infty$. Numerical calculations tell us that
$0<\kappa\le 1$ for the regions of interest in the runaway tail, and
it is often slowly varying function of $\spa$. That is, the
characteristic width of the distribution function in the $\spe$
direction, $1/W^2$, decreases with increasing $\spa$, and slowly
asymptotes to a constant value. For now, we may simply use $0 <
\kappa\le 1$ as a working hypothesis to be verified through numerical
calculations later.
  
We start with the Eq.~(\ref{eq:extremaequation}) satisfied at extrema or inflection of the
distribution function and rewrite it as
\begin{equation}
L(\spa)\equiv 2\left(\sigma \spa +\sqrt{1+\spa^2}\right)-K(\spa)(1+\spa^2)=0,
\label{eq:extremaeq2}
\end{equation}
where $K(\spa)=\Winf^2 (1+\Zeff) \kappa (\spa)=\Ebarm \sigma
\kappa(\spa)$ with $\Ebar=(\Eh-1)/[2(1+\Zeff)]$. It is useful to form
\begin{equation}
L'(\spa)/2\approx \sigma  +\frac{\spa}{\sqrt{1+\spa^2}}-K(\spa)\spa,
\label{eq:extremaeqprime}
\end{equation}
where prime denotes a derivative with respect to $\spa$, and we
neglected a term, $-K'(\spa)(1+\spa^2)/2 $, assuming $\kappa$ to be
sufficiently slowly varying function of $\spa$. From
Eq.~(\ref{eq:extremaeqprime}) we see that $L'(\spa)\rightarrow
-\infty$ as $\spa\rightarrow \infty$, while $L'|_{\spa=0}=2\sigma >
0$. It can also be shown that $L'=0$ only has one root for positive
values of $\spa$, which then has to correspond to a single maximum of
$L$.

If the distribution has an inflection point, both $L$ and $L'$ should
vanish there. Assuming the slowly varying $\kappa$ to be a constant,
the system of equations (\ref{eq:extremaeq2}) and
(\ref{eq:extremaeqprime}) can be solved for $\spa$ and $K$ to find
\begin{align}
\label{eq:ksp1}
K_0\equiv K(\spz)=
&(4\sqrt{2}\sigma)^{-1}\left[8\sigma\sqrt{4-\frac{8}{3+\sqrt{1+8\sigma^2}}}\right.
  \\ +&\left.\left(1+4\sigma^2-\sqrt{1+8\sigma^2} \right)\sqrt{2
    -\frac{8}{3+\sqrt{1+8\sigma^2}}} \right]\nonumber,
\end{align}
and
\begin{align}
\label{eq:spz1}
\spz=\sqrt{1-\frac{4}{3+\sqrt{1+8\sigma^2}}},
\end{align}
where the subscript $0$ refers to values of quantities at the
threshold of the bump appearance.  By inspecting the expressions
(\ref{eq:ksp1}) and (\ref{eq:spz1}) we find that both of them increase
with $\sigma$ monotonically; $K_0$ between $2$ and $\infty$, and
$\spz$ between $0$ and $1$. This means that an inflection point is
always located below $\spa=1$. Note, that we assume the inflection
point to be sufficiently far from the bulk, and that $\kappa'$ is
small; violation of these assumptions may move $\spz$ above
unity. However, since this problem cannot be addressed until the more
complete Eq.~(\ref{eq:final_kinetic_equation}) is solved, we assume
that the conditions above are fulfilled in order to proceed
analytically. Since $s_{\|}<1$, we can make use of the expansion
\begin{align}
\label{eq:expandroot}
\sqrt{1+s_{\parallel}^2} = 1+\frac{s_{\parallel}^2}{2}+\mathcal{O}(s_{\parallel}^4).
\end{align}  
By neglecting $\mathcal{O}(s_{\parallel}^4)$ terms (which is
reasonably good even when $\spa$ approaches unity),
Eq.~(\ref{eq:extremaeq2}) becomes quadratic in $\spa$:
\begin{equation}
2(\sigma \spa +1+\spa^2/2)-K(\spa)(1+\spa^2)=0.
\label{eq:extremaexp}
\end{equation}
 At the inflection point $\spz$, Eq.~(\ref{eq:extremaexp}) must have a
 single root, which requires the discriminant to vanish. This
 determines the threshold value of $K$
\begin{equation}
\label{eq:kapprox}
K_0=\left(3+\sqrt{1+4\sigma^2} \right)/2, 
\end{equation}
which is positive. This can be substituted back into
Eq.~(\ref{eq:extremaexp}) to find
\begin{equation}
\label{eq}
\spz=\sigma/(K_0-1)=\sigma \left[\left(1+\sqrt{1+4\sigma^2} \right)/2
  \right]^{-1}.
\end{equation}
Combining $K(\spa)=\Ebarm \sigma \kappa(\spa)$ with
Eq.~(\ref{eq:kapprox}) to solve for a positive $\sigma$ that corresponds
to $K\ge 2$, yields $\sigma$ as a function of $\Ebar$ at the
threshold for bump formation
\begin{equation}
\label{eq:thresholdcondition}
\sigma_0=\frac{3\kappa/\Ebar+\sqrt{8+\kappa^2/\Ebar^2}}{2\left(\kappa^2/\Ebar^2-1 \right)},
\end{equation}
where $\kappa=\kappa(\spa=\spz)\le 1$ is treated as a parameter.  When
$\sigma$ is increased above $\sigma_0$, $L$ becomes negative, and no
bump appears. Reducing $\kappa$ below unity increases the threshold
value of $\sigma$. Thus Eq.~(\ref{eq:thresholdcondition}) with
$\kappa=1$ represents an absolute lower threshold in $\sigma$ for a
monotonic behavior of the steady state distribution
function. Furthermore, even at $\kappa=1$ the threshold is limited to
the region $\Ebar< 1$, thus a bump should always appear when $\Ebar\ge
1$.

We have considered $K(\spa=0)>2$, in which case $L(\spa)=0$ can have
zero (no bump), one (threshold), or two (bump exists) positive real
roots. When $K(\spa=0)<2$ there is only a single positive root of
$L(\spa)$. Since the distribution function cannot have positive slope
in the high $\spa$ limit this root should also correspond to a
bump. That, however, requires the existence of a minimum in the
distribution function along the positive $\spa$ axis, which must then
appear outside the domain of validity of
Eq.~(\ref{eq:extremaequation}). In fact, this minimum will appear
close to the bulk part of the electron distribution, where neglected
corrections to the collision operator become important.

We can conclude that for $\Ebar\ge 1$, there should always be a bump
in the \emph{steady state} distribution function as long as there is a
finite magnetic field. This, perhaps somewhat counter-intuitive,
result needs some clarification to accommodate the well known
$\sigma=0$ limit. When no loss mechanisms are considered (in this case when
$\sigma=0$), the electron distribution has no steady state solution,
and the runaway tail at $\spe=0$ should converge to a $1/\spa$
decay. When $\sigma$ is small, the bump location moves to high values
in $\spa$, as will be shown in the next section. The runaway tail
always builds up starting from the bulk and, for a tiny $\sigma$, the
process may take such a long time that the distribution never becomes
non-monotonic in practice. In this scenario the steady state
distribution and the bump have no relevance. Also, other loss
mechanisms may limit the distribution function to momenta below $\spz$
in realistic cases.

\subsection{An estimate for the bump location in the far-tail}
In order to proceed and estimate the location of the bump and the
shape of the distribution function, we look
for a steady-state solution in a region where the
guiding-center parallel momentum is large. Using the expansion
 \begin{align}
\sqrt{1+s_{\parallel}^2}=s_{\parallel}+\mathcal{O}(s_{\parallel}^{-1}),
\end{align}
Eq.~(\ref{eq:accurate_limit}) gives
\begin{equation}
\label{eq:approximate_kinetic_equation}
\left(\hat{E}-1+\mathcal{O}(s_{\parallel}^{-1})\right)\left[\frac{\partial F}{\partial s_{\parallel}}\right]_{(s_{\parallel},0)}-(1+Z_{\textrm{eff}})\left[\frac{\partial^{2} F}{\partial s_{\perp}^{2}}\right]_{(s_{\parallel},0)}-2\frac{1+\sigma}{s_{\parallel}} \left[F\right]_{(s_{\parallel},0)}=0.
\end{equation}
We neglect the $\mathcal{O}(s_{\parallel}^{-1})$-term, which is valid
if $\hat{E}-1$ is not very small, and assume that the width of the
distribution function in the $s_{\perp}$-direction, and thus
$W^2=-F^{-1}(\partial^2 F/\partial \spe^2)|_{(\spa,0)}$, stays
approximately constant close to the bump.  This essentially means that
we are looking for a separable solution of the form $F\sim
h(s_{\parallel})g(s_{\perp})$. We obtain an ordinary differential
equation
\begin{align}
\left(\hat{E}-1\right)s_{\parallel}h'-2(1+\sigma)h=-(1+Z_{\textrm{eff}})W^2s_{\parallel}h,
\end{align}
which is solved by
\begin{align}
h(s_{\parallel})\sim s_{\parallel}^{2(1+\sigma)/(\hat{E}-1)}\exp{\left[-W^2(1+Z_{\textrm{eff}})/(\hat{E}-1)s_{\parallel}\right]},
\end{align}
and the location of the bump-on-tail is given by
\begin{align}
s_{\parallel}=\frac{1+\sigma}{1+Z_{\textrm{eff}}}\frac{2}{\Winf^2\kappa}.
\label{eq:location1}
\end{align}
If we again assume that $\kappa$ does not exceed unity, recalling $\Winf^2=2\sigma/(\hat{E}-1)$, we find a lower bound for the
parallel momentum at the bump
\begin{align}
\label{eq:location}
s_{\parallel,\min}=\frac{1+\sigma}{\sigma}\frac{\hat{E}-1}{1+Z_{\textrm{eff}}}=\frac{1+\sigma}{\sigma}2\Ebar.
\end{align}
We see that for small values of $\sigma$, the bump would appear at
high parallel momenta. By setting $s_{\parallel,\min}$ to some upper
limit of physical interest, $s_{\|,\rm L}$,
Eq.~(\ref{eq:location}) may be used to find an estimate for a
lower ``practical limit'' in $\sigma$ for the appearance of the
bump. Namely, if $\sigma$ is smaller than
\begin{equation}
\label{eq:practicallimit}
\sigma_{\rm L}= \frac{1}{(s_{\|,\rm L}\kappa)
  /(2\Ebar)-1},
\end{equation}
for $\kappa=1$, then a bump would only appear at some large parallel
momentum above $s_{\|,\rm L}$, which is then deemed physically
irrelevant.  Note that if the bump is in the far tail, $\kappa$ can be
significantly less than unity, as will be shown in the next section,
using numerical simulations. Letting $\kappa<1 $ increases the
practical limit in $\sigma$. Another implication of
Eq.~(\ref{eq:practicallimit}) is that for a normalized electric field
higher than $\Ebar=s_{\|,\rm L}/2$, the bump \emph{always} appears
above $s_{\|,\rm L}$ for any value of $\sigma$.
 
\section{Comparison to numerical results}
\label{sec:numres}
The numerical results shown in this section were performed with the
continuum simulation tool, CODE, used in its time-independent mode.
CODE solves the two dimensional momentum space kinetic equation in a
homogeneous plasma, using a linearized Fokker-Planck operator valid
for arbitrary electron energies. For a detailed description of
the tool, see \citep{Landreman2014847}.

\begin{figure}
\begin{center}
\includegraphics[width=0.5\textwidth]{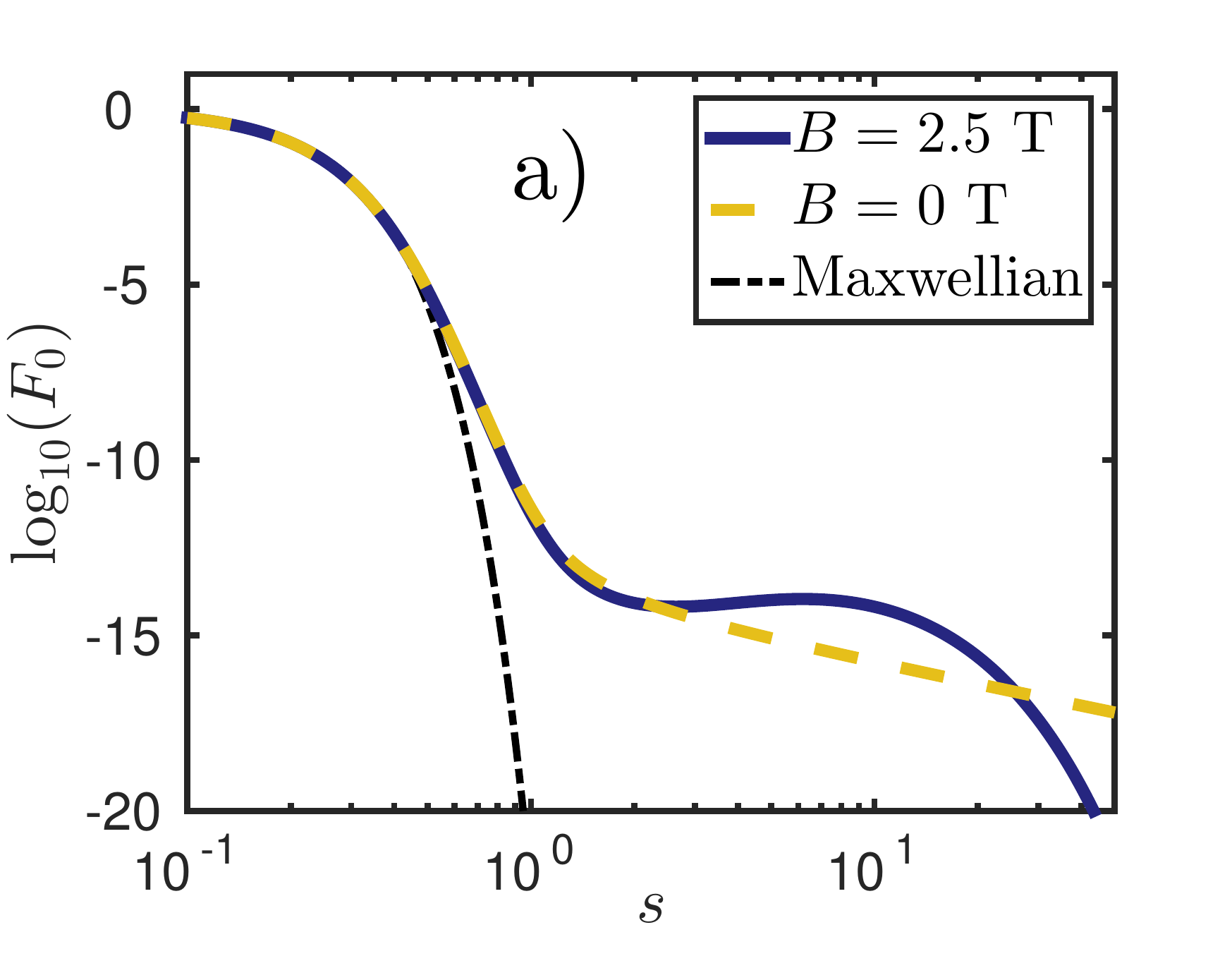}
\hspace{-5.5mm}
\includegraphics[width=1\textwidth]{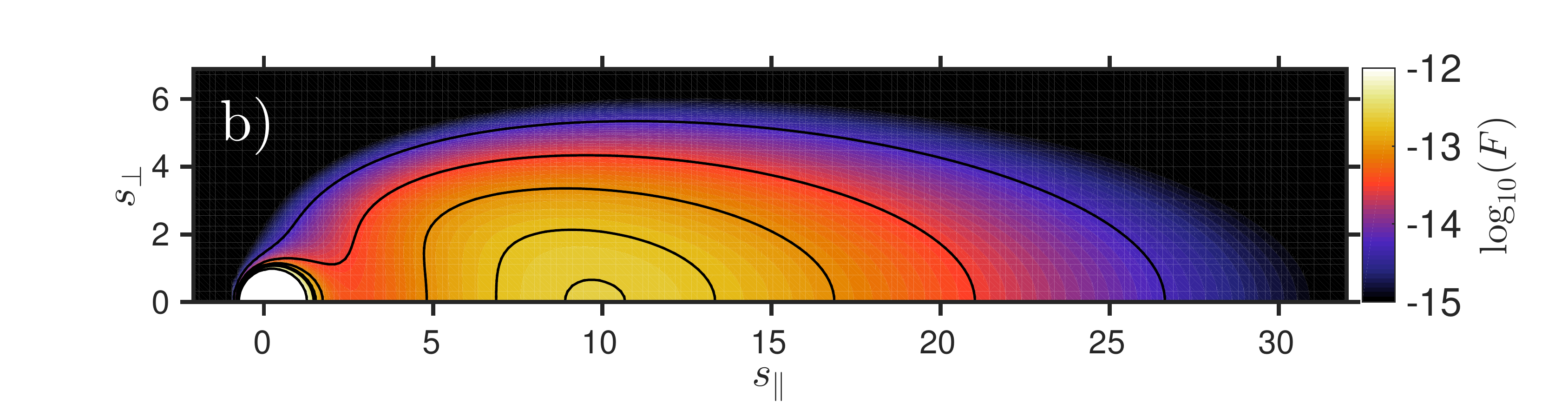}
\caption{Typical examples of non-monotonic runaway electron
  distribution functions. a) The pitch angle average of the
  distribution function with (solid curve) and without (dashed)
  synchrotron radiation reaction. A Maxwellian distribution is also
  indicated (dash-dotted). b) Contour plot of the distribution
  function corresponding to the solid curve in (a), as a
  function of $\spa$ and $\spe$.}
\label{fig:typicaldistribution}
\end{center}
\end{figure}

First, we provide a typical example of a non-monotonic runaway
distribution function in the presence of radiation
reaction. Figure~\ref{fig:typicaldistribution}a) shows the momentum
dependence of the pitch angle averaged runaway electron distribution
with ($B=2.5\,\rm{T}$) and without ($B=0\,\rm{T}$) radiation reaction
force, plotted with solid and dashed curves respectively. Technically,
the pitch-angle-averaged distribution is the lowest mode in a Legendre
polynomial expansion of $F$ in $\xi$, normalized so that $F$ is unity
at its maximum. The simulations were performed with the parameters
$T_e=5\,\textrm{keV}$, $n_e=2\cdot 10^{19}\,\mathrm{m}^{-3}$,
$\Zeff=1.2$, and $\Eh=2$. Note that the distribution function without
radiation reaction represents a quasi-steady state.  The lack of loss
mechanisms leads to a slow but steady depletion of the bulk electron
population, as more and more electrons run away and leave the
computational domain. This outflow must be balanced by an artificial
source of thermal (Maxwellian) electrons to maintain the quasi-steady
state. In the presence of radiation reaction, the distribution is a
true steady state. When the radiation reaction is included, the
non-monotonic feature is present when the distribution is averaged
over pitch angles in the present example. However, we note that for
less pronounced bumps, the pitch-angle-averaged distribution can have
a monotonic tail, or it may exhibit a bump at some $s$ which are
appreciably lower than those where the bump is observed in the full
2D-distribution. This may have an impact on the possibility of
bump-on-tail type instabilities to arise.

Figure~\ref{fig:typicaldistribution}b) shows a contour plot in
$\spa$--$\spe$ momentum space of the distribution function
corresponding to the solid curve in
Fig.~\ref{fig:typicaldistribution}a). Although this example is
representative of a typical runaway electron distribution, the
location and the height of the bump, and the width of the distribution
in $\spe$ can vary significantly depending on the plasma
parameters. The relation between the location $\spa$ and the local
``width'' [$\sim 1/W^2=1/(\Winf^2\kappa)$] of the distribution given
by Eq.~(\ref{eq:location1}) is accurate as long as the location of the
bump is not close to unity, i.e.  sufficiently far from the no-bump
threshold (\ref{eq:thresholdcondition}). This justifies the
approximations applied to the collision operator in our analysis. In
particular, energy diffusion can be neglected since no sharp features
of the distribution function in $\spa$ are present, as seen in
Figure~\ref{fig:typicaldistribution}b).

In order to investigate the validity of our analytical calculations,
we have performed a numerical analysis of the appearance of the bump
by scanning the parameter space with CODE.  The electron temperature
and density where held constant at the values $T_e=1\,\textrm{keV}$
and $n_e=5\cdot 10^{18}\,\mathrm{m}^{-3}$, respectively, while the
magnetic field, the induced electric field and the effective ion
charge were varied over the ranges $B\in [1,6]\,\rm T$, $\Eh \in
[2,14]$ and $\Zeff\in [1,3]$. The numerical calculations used $950$
momentum grid points, $130$ Legendre modes for the decomposition in
$\xi$, and a highest resolved momentum of $s=34$, provided well
converged solutions.

\begin{figure}
\begin{center}
\centerline{\includegraphics[width=0.65\textwidth]{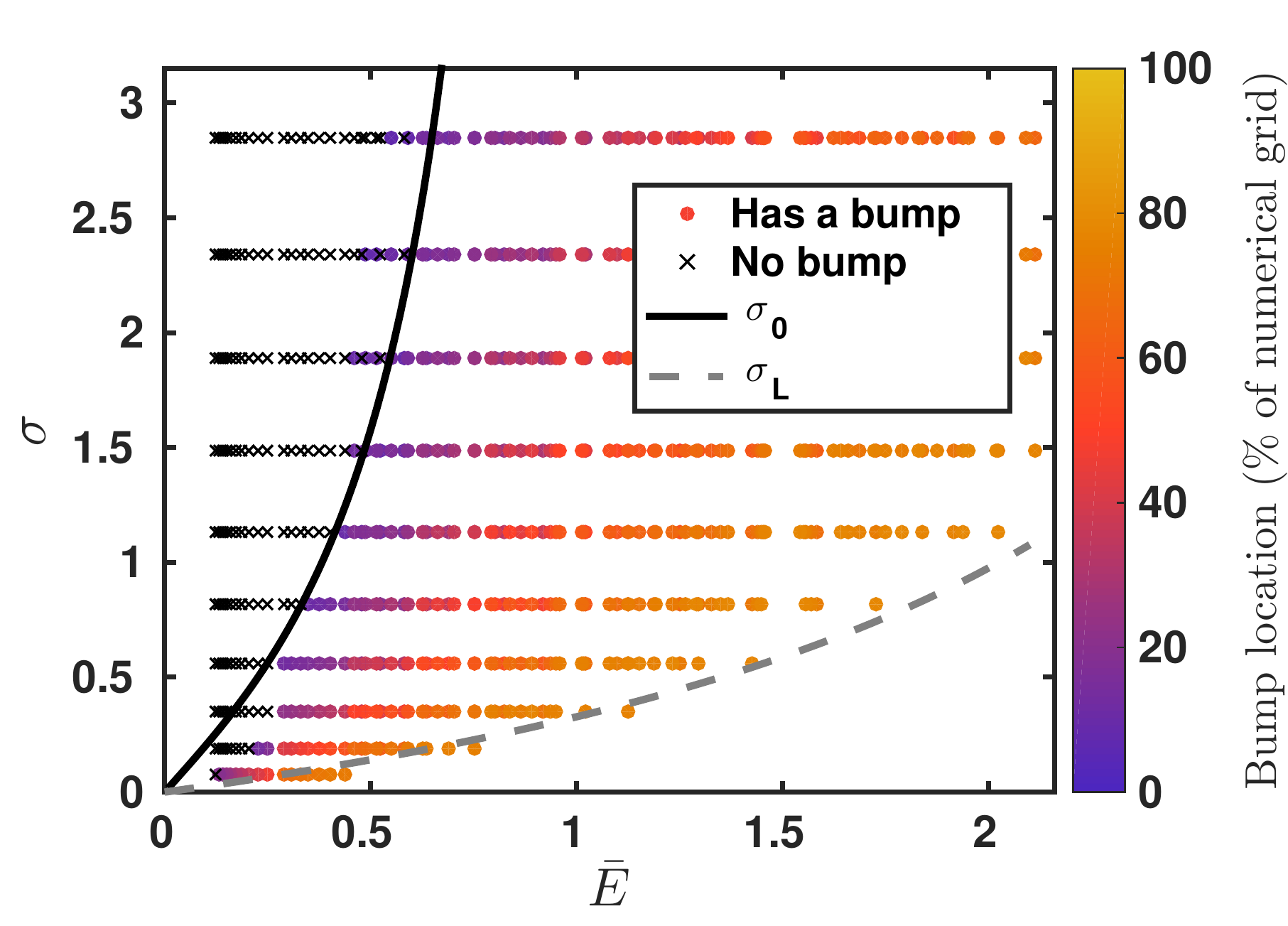}}
\caption{Parameter scan of CODE simulations yielding steady
  state solutions with (colored circles) or without (black crosses) a bump
  in the runaway tail. Good correlation is found with the analytical
  threshold condition given in Eq.~(\ref{eq:thresholdcondition}) for
  $\kappa=1$ (solid line). The dashed line
  represents the ``practical threshold'' in Eq.~\eqref{eq:practicallimit} for 
  \mbox{$s_{\|,\rm L}=27$} and $\kappa=0.3$. The color coding shows the 
  location of the bump relative to $s_{\|}=34$.  }
\label{fig:appearance}
\end{center}
\end{figure}

The results of the scan are presented in
Fig.~\ref{fig:appearance}, where circles and crosses correspond
to distributions with and without a bump, respectively. The color coding of
the circles reflects the location of the bump, with $100\%$ in
the color bar corresponding to $\spa=34$. Simulations with a bump
appearing above $80\%$ ($\spa=27$) are excluded from the figure, since
those results may be affected by the bump being too close to the
highest resolved momentum. As expected from Eq.~(\ref{eq:location}),
increasing $\Ebar$ or decreasing $\sigma$ moves the bump towards larger
momenta.

A reasonably good agreement is found between the numerical
calculations and the analytical threshold for the bump to exist. The
solid curve shows this threshold, Eq.~(\ref{eq:thresholdcondition}),
for $\kappa=1$.  Above $\sigma\approx 0.5$ the ``no bump'' solutions
obey the analytical threshold and fall to the left of the threshold
curve. There are some solutions with bump in this region as well,
however this is not surprising, since $\kappa$ at the bump is allowed
to be less than unity, in which case the threshold moves towards lower
values of $\Ebar$. The threshold begins to fail for lower values of
$\sigma$, showing that the approximation $K'\ll \{\sigma, \spa\}$ used
in Eq.~(\ref{eq:extremaeqprime}) breaks down. Nevertheless, the
qualitative behavior of the threshold is still captured by
Eq.~(\ref{eq:thresholdcondition}). The lower right corner of the plot
(high $\Ebar$ and small $\sigma$) is not populated, since
some simulations where the bump would have appeared at too high
$\spa$ were excluded. With $s_{\|,\rm L}=27$, a $\kappa$ value as low
as $0.3$ is needed in order for the limit given by
Eq.~(\ref{eq:practicallimit}) (dashed line) to correspond well to 
the boundary of the region of excluded points. Thus, $\kappa$ can be 
significantly lower than unity at a bump with large momentum.
 
\begin{figure}
\begin{center}
\includegraphics[width=0.6\textwidth]{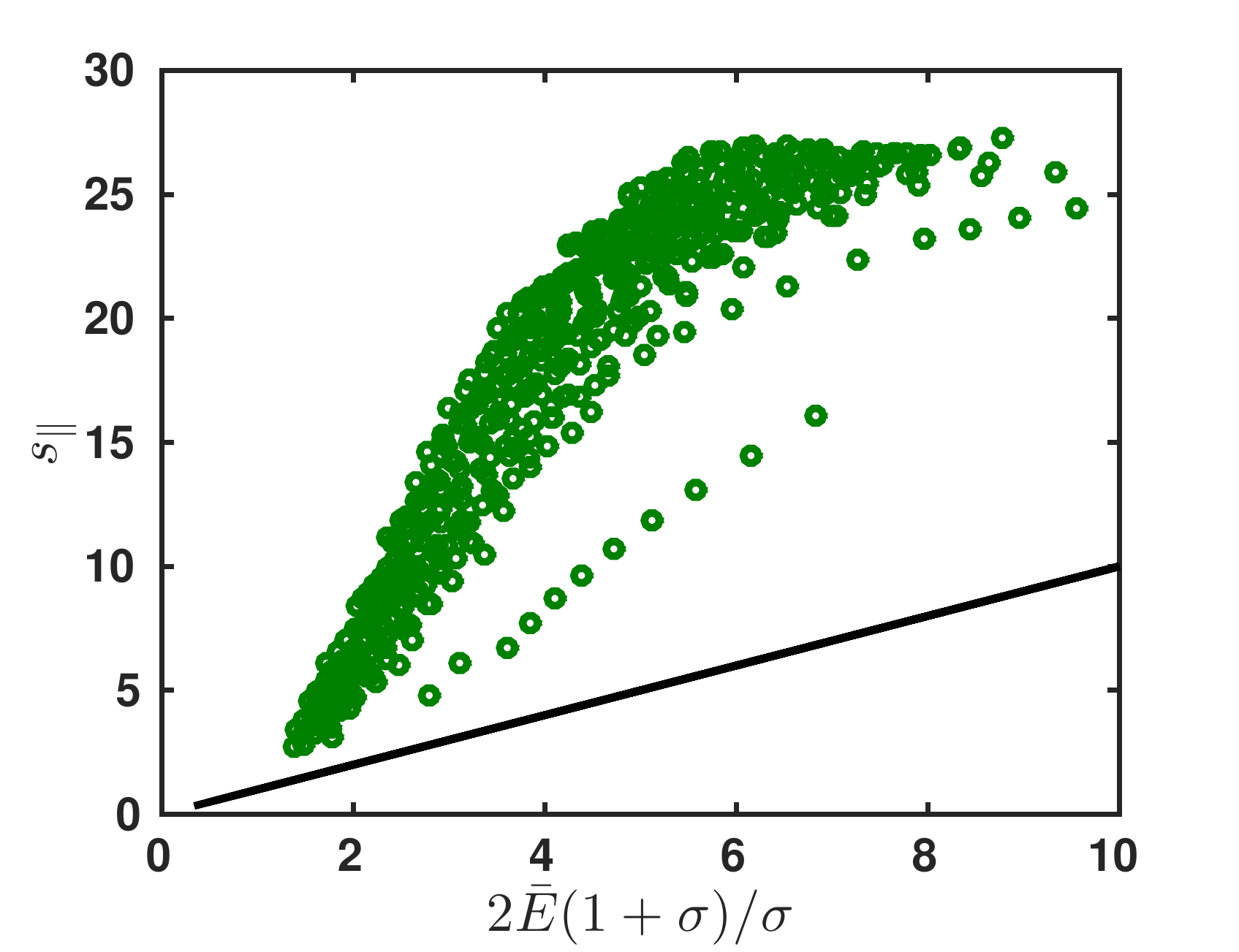}
\caption{Parallel momentum of the bump. Circles denote the
  locations of the bumps according to numerical solutions, while the
  solid line represents a theoretical lower limit, 
  Eq.~(\ref{eq:location}). In the simulations, all bumps appear above 
  the analytical threshold condition.}
\label{fig:location}
\end{center}
\end{figure}

From the parameter scan used to generate Fig.~\ref{fig:appearance}, the 
simulations exhibiting bumps were compared to the theoretical lower bound 
for the location of the bump (Eq.~\ref{eq:location}). As shown in Fig.~\ref{fig:location}, we find
that, indeed, the parallel momenta at the bumps (shown with green
circles) are all higher than the lower bound (solid line). In fact,
most of the $\spa$ values are well above this limit. This merely
confirms our finding that $\kappa$ at the bump is typically less than
unity, especially for parameters sufficiently exceeding the no-bump
threshold.

\section{Conclusions}
\label{sec:conclusions}
We have analyzed the runaway electron distribution function,
accounting for the radiation reaction force, and have 
shown that the steady state runaway
distribution can become non-monotonic. Furthermore, a threshold
condition for the appearance of the bump, as well as a lower limit to its
location in momentum space, were derived. While slowing down and
pitch-angle scattering due to Coulomb collisions are taken into
account in our analysis, we do not consider the effect of large angle
collisions, and we restrict the study to a straight magnetic field
geometry. Our analytical results show good agreement with
numerical simulations obtained using the CODE solver.

We find that for a normalized electric field larger than unity,
  $\Ebar>1$, the steady state electron distribution \emph{always}
  exhibits a bump, independently of the value of the $\sigma$
  parameter quantifying the strength of the radiation reaction, as
  long as the magnetic field is non-vanishing ($\sigma>0$). For a
smaller electric field, the appearance of the bump is well correlated
with the $\sigma$ threshold given by
Eq.~(\ref{eq:thresholdcondition}). Although above this threshold there
must always be a bump in the steady state distribution function, it
may not have a practical relevance in some cases.  When $\sigma$ is
small and/or $\Ebar$ is large, the bump would be located at very large
parallel momentum, but the forefront of the electron distribution can
require a long time to reach that far. This motivates the introduction
of another ``practical'' threshold condition,
Eq.~(\ref{eq:practicallimit}).  If $\sigma$ is lower than this
threshold, the bump will appear at a momentum above some specified
limit, $s_{\|,\rm L}$, and can then be considered unimportant. In
particular, above a normalized electric field of $\Ebar=s_{\|,\rm
  L}/2$, this criterion is satisfied for any $\sigma$.

Nevertheless, when the radiation reaction is strong enough and/or the
parallel electric field is not too high, there is a possibility for a
bump to form in the runaway tail. This non-monotonic feature presents
a potential source for bump-on-tail instabilities, which can play a
role in limiting the formation of large runaway beams.

The authors are grateful to M. Landreman and P. Helander for fruitful
discussions. IP was supported by the International Postdoc grant of
Vetenskapsr{\aa}det.

\appendix
\section{The relativistic collision operator}\label{app:relativistic_collision_operator}
The relativistic particle phase-space Beliaev-Budker Collision operator in the Landau form is defined as
\begin{align}
C[f_a,f_b]=-\frac{\Gamma_{ab}}{2}\frac{\partial}{\partial\mathbf{p}}\cdot\int d\mathbf{p}'\mathbb{U}(\mathbf{u},\mathbf{u}')\cdot\left(f_a\frac{\partial f_b}{\partial\mathbf{p}'}-f_b\frac{\partial f_a}{\partial\mathbf{p}}\right),
\end{align}
where $\Gamma_{ab}=e_a^2e_b^2\ln\Lambda/(4\pi\varepsilon_0^2)$, $\mathbf{u}=\mathbf{p}/m_a$, $\mathbf{u}'=\mathbf{p}'/m_b$, and the collision kernel is given by
\begin{align}
\mathbb{U}(\mathbf{u},\mathbf{u}')=\frac{r^2}{\gamma\gamma'w^3}\left[w^2\mathbb{I}-\mathbf{u}\mathbf{u}-\mathbf{u}'\mathbf{u}'+r(\mathbf{u}\mathbf{u}'+\mathbf{u}'\mathbf{u})\right],
\end{align}
with the coefficients
\begin{align}
\gamma=&\sqrt{1+(u/c)^2},\\
\gamma'=&\sqrt{1+(u'/c)^2},\\
r=&\gamma\gamma'-\mathbf{u}\cdot\mathbf{u}'/c^2,\\
w=&c\sqrt{r^2-1}.
\end{align}
\citet{braams_karney:1989} found a corresponding differential form for the collision operator
\begin{align}
C[f_a,f_b]=-\frac{\partial}{\partial\mathbf{p}}\cdot\left(\mathbf{K}_{ab}[f_b]f_a-\mathbb{D}_{ab}[f_b]\cdot\frac{\partial f_a}{\partial\mathbf{p}}\right),
\end{align}
where $\mathbf{K}_{ab}[f_b]$ is the friction vector and $\mathbb{D}_{ab}[f_b]$ the diffusion tensor that are defined with differential operations on Braams--Karney potentials $\Psi_n(u)$ according to
\begin{align}
\mathbf{K}_{ab}[f_b]=&-4\pi\frac{m_a}{m_b}\frac{\Gamma_{ab}}{\gamma}\mathsf{K}\left(\Psi_1-2\frac{\Psi_2}{c^2}\right),\\
\mathbb{D}_{ab}[f_b]=&-4\pi\frac{\Gamma_{ab}}{\gamma}\left[\mathsf{L}\left(\Psi_3-4\frac{\Psi_4}{c^2}\right)+\frac{1}{c^2}\left(\mathbb{I}+\frac{\mathbf{u}\mathbf{u}}{c^2}\right)\left(\Psi_3+4\frac{\Psi_4}{c^2}\right)\right].
\end{align}

The differential operators $\mathsf{K}$ and $\mathsf{L}$ are defined as 
\begin{align}
\mathsf{K}\Psi(u)=&\left(\mathbb{I}+\frac{\mathbf{u}\mathbf{u}}{c^2}\right)\cdot\frac{\partial\Psi}{\partial\mathbf{u}},\\
\mathsf{L}\Psi(u)=&\left(\mathbb{I}+\frac{\mathbf{u}\mathbf{u}}{c^2}\right)\cdot\frac{\partial^2\Psi}{\partial\mathbf{u}\partial\mathbf{u}}\cdot\left(\mathbb{I}+\frac{\mathbf{u}\mathbf{u}}{c^2}\right)+\frac{1}{c^2}\left(\mathbb{I}+\frac{\mathbf{u}\mathbf{u}}{c^2}\right)\left(\mathbf{u}\cdot\frac{\partial\Psi}{\partial\mathbf{u}}\right)
\end{align}
and the potential functions are given by the integrals
\begin{align}
\Psi_0(\mathbf{u})=&-\frac{1}{4\pi}\int d\mathbf{u}'\frac{f_b(\mathbf{u}')}{\gamma'w},\\
\Psi_1(\mathbf{u})=&-\frac{1}{4\pi}\int d\mathbf{u}'\frac{r f_b(\mathbf{u}')}{\gamma'w},\\
\Psi_2(\mathbf{u})=&-\frac{1}{8\pi}\int d\mathbf{u}'\sinh^{-1}(w/c)\frac{c\, f_b(\mathbf{u}')}{\gamma'},\\
\Psi_3(\mathbf{u})=&-\frac{1}{8\pi}\int d\mathbf{u}'\frac{w\, f_b(\mathbf{u}')}{\gamma'},\\
\Psi_4(\mathbf{u})=&-\frac{1}{32\pi}\int d\mathbf{u}'\frac{c^3}{\gamma}\left(r\sinh^{-1}(w/c)-(w/c)\right)f_b(\mathbf{u}').
\end{align} 
Furthermore, the potential functions satisfy differential relations
\begin{align}
L_0\Psi_0=&\;f_b,\\
L_1\Psi_1=&\;f_b,\\
L_1\Psi_2=&\;\Psi_1,\\
L_2\Psi_3=&\;\Psi_0,\\
L_2\Psi_4=&\;\Psi_3,
\end{align}
where the operator $L_k$ is defined
\begin{align}
L_k\Psi=\left(\mathbb{I}+\frac{\mathbf{u}\mathbf{u}}{c^2}\right):\frac{\partial^2\Psi}{\partial\mathbf{u}\partial\mathbf{u}}+\frac{3\mathbf{u}}{c^2}\cdot\frac{\partial\Psi}{\partial\mathbf{u}}+\frac{1-k^2}{c^2}\Psi.
\end{align}
In the case of isotropic background distributions $f_b(u)$, the
potentials become functions of $u$ only, and the friction vector and
diffusion tensor can be simplified into
\begin{align}
\mathbf{K}_{ab}=&-4\pi\frac{m_a}{m_b}\Gamma_{ab}\gamma\left(\frac{\partial\Psi_1}{\partial u}-\frac{2}{c^2}\frac{\partial\Psi_2}{\partial u}\right)\frac{\mathbf{p}}{p}\equiv -\nu_{l,ab}\mathbf{p},\\
\mathsf{D}_{ab}=&-4\pi\Gamma_{ab}\gamma\left(\Psi_0-\frac{2\gamma^2}{u}\frac{\partial\Psi_3}{\partial u}+\frac{8\gamma^2}{uc^2}\frac{\partial\Psi_4}{\partial u}-\frac{8}{c^4}\Psi_4\right)\frac{\mathbf{p}\mathbf{p}}{p^2}\nonumber\\
&-4\pi\Gamma_{ab}\gamma\left(\frac{1}{u}\frac{\partial\Psi_3}{\partial u}+\frac{1}{c^2}\Psi_3-\frac{4}{uc^2}\frac{\partial\Psi_4}{\partial u}+\frac{4}{c^4}\Psi_4\right)\left(\mathbb{I}-\frac{\mathbf{p}\mathbf{p}}{p^2}\right)\nonumber\\
\equiv&\;D_{l,ab}\frac{\mathbf{p}\mathbf{p}}{p^2}+D_{t,ab}\left(\mathbb{I}-\frac{\mathbf{p}\mathbf{p}}{p^2}\right).
\end{align}

The guiding-center transformation of the Fokker-Planck operator
presented in~\citep{brizard:4429} gave explicit expressions for the
guiding-center friction and diffusion coefficients,
$\left\langle\mathcal{K}_{ab,gc}^{\alpha}\right\rangle$ and
$\left\langle\mathcal{D}_{ab,gc}^{\alpha\beta}\right\rangle$, in the
case of isotropic background distributions and a non-relativistic
collision kernel. Generalization of that work to relativistic
collision kernel is straight-forward in the case of isotropic
field-particle distributions because the forms of the particle
phase-space friction and diffusion coefficients do not change. Only
the expressions for $\nu_{l,ab}$, $D_{l,ab}$, and $D_{t,ab}$ are
different but that will not affect the guiding-center transformation,
as they are functions only of the guiding-center kinetic momentum.

\bibliographystyle{jpp}
\bibliography{bump_jpp}

\providecommand{\noopsort}[1]{}\providecommand{\singleletter}[1]{#1}
\begin{thebibliography}{20}
\expandafter\ifx\csname natexlab\endcsname\relax\def\natexlab#1{#1}\fi

\bibitem[Abraham(1905)]{abraham}
{\sc Abraham, M.} 1905 {\em Theorie der Elektrizit\"{a}t, Vol II:
  Elektromagnetische Theorie der Strahlung\/}. Teubner Leipzig.

\bibitem[Andersson {\em et~al.\/}(2001)Andersson, Helander \&
  Eriksson]{andersson:pop2001}
{\sc Andersson, F., Helander, P. \& Eriksson, L.-G.} 2001 Damping of
  relativistic electron beams by synchrotron radiation. {\em Physics of
  Plasmas\/} {\bf 8}~(12), 5221--5229.

\bibitem[Braams \& Karney(1989)]{braams_karney:1989}
{\sc Braams, Bastiaan~J. \& Karney, Charles F.~F.} 1989 Conductivity of a
  relativistic plasma. {\em Physics of Fluids B\/} {\bf 1}~(7), 1355--1368.

\bibitem[Brizard(2004)]{brizard:4429}
{\sc Brizard, A.~J.} 2004 A guiding-center {Fokker}--{Planck} collision
  operator for nonuniform magnetic fields. {\em Physics of Plasmas\/} {\bf
  11}~(9), 4429--4438.

\bibitem[Cary \& Brizard(2009)]{RevModPhys.81.693}
{\sc Cary, John~R. \& Brizard, Alain~J.} 2009 Hamiltonian theory of
  guiding-center motion. {\em Reviews of Modern Physics\/} {\bf 81}, 693--738.

\bibitem[Decker {\em et~al.\/}(2010)Decker, Peysson, Brizard \&
  Duthoit]{decker:112513}
{\sc Decker, J., Peysson, Y., Brizard, A.~J. \& Duthoit, F.-X.} 2010
  Orbit-averaged guiding-center {F}okker-{P}lanck operator for numerical
  applications. {\em Physics of Plasmas\/} {\bf 17}~(11), 112513.

\bibitem[Di~Piazza {\em et~al.\/}(2012)Di~Piazza, M\"uller, Hatsagortsyan \&
  Keitel]{RevModPhys.84.1177}
{\sc Di~Piazza, A., M\"uller, C., Hatsagortsyan, K.~Z. \& Keitel, C.~H.} 2012
  Extremely high-intensity laser interactions with fundamental quantum systems.
  {\em Rev. Mod. Phys.\/} {\bf 84}, 1177--1228.

\bibitem[Dirac(1938)]{dirac:1938}
{\sc Dirac, P. A.~M.} 1938 Classical theory of radiating electrons. {\em
  Proceedings of the Royal Society of London. Series A, Mathematical and
  Physical Sciences\/} {\bf 167}~(929), pp. 148--169.

\bibitem[Ford \& O'Connell(1993)]{Ford1993182}
{\sc Ford, G.W \& O'Connell, R.F} 1993 Relativistic form of radiation reaction.
  {\em Physics Letters A\/} {\bf 174}~(3), 182 -- 184.

\bibitem[Hazeltine \& Mahajan(2004)]{hazeltine2004}
{\sc Hazeltine, R. \& Mahajan, S.} 2004 Radiation reaction in fusion plasmas.
  {\em Physical Review E\/} {\bf 70}, 046407.

\bibitem[Hirvijoki {\em et~al.\/}(2013)Hirvijoki, Brizard, Snicker \&
  Kurki-Suonio]{hirvijoki:092505}
{\sc Hirvijoki, E., Brizard, A., Snicker, A. \& Kurki-Suonio, T.} 2013 Monte
  {C}arlo implementation of a guiding-center {F}okker-{P}lanck kinetic
  equation. {\em Physics of Plasmas\/} {\bf 20}~(9), 092505.

\bibitem[Hirvijoki {\em et~al.\/}(2015)Hirvijoki, Decker, Brizard \&
  Embreus]{hirvijoki:2014}
{\sc Hirvijoki, E., Decker, J., Brizard, A. \& Embreus, O.} 2015 Guiding-center
  transformation of the {A}braham-{L}orentz-{D}irac radiation reaction force.
  {\em Submitted to Journal of Plasma Physics\/} .

\bibitem[Landau \& Lifshitz(1975)]{landau_lifshitz_fields}
{\sc Landau, L.~D. \& Lifshitz, E.~M.} 1975 {\em The Classical Theory of
  Fields\/}, fourth edition edn., {\em Course of Theoretical Physics\/},
  vol.~2. Amsterdam: Pergamon.

\bibitem[Landreman {\em et~al.\/}(2014)Landreman, Stahl \&
  F{\"u}l{\"o}p]{Landreman2014847}
{\sc Landreman, Matt, Stahl, Adam \& F{\"u}l{\"o}p, T{\"u}nde} 2014 Numerical
  calculation of the runaway electron distribution function and associated
  synchrotron emission. {\em Computer Physics Communications\/} {\bf 185}~(3),
  847 -- 855.

\bibitem[Littlejohn(1983)]{littlejohn:jpp:4732464}
{\sc Littlejohn, Robert~G.} 1983 Variational principles of guiding centre
  motion. {\em Journal of Plasma Physics\/} {\bf 29}, 111--125.

\bibitem[Lorentz(1892)]{lorentz2}
{\sc Lorentz, H.A.} 1892 La {T}h\'{e}orie \'{E}lectromagn\'{e}tique de
  {Maxwell} et {S}on {Application} {Aux} {Corps} {Mouvants}. {\em Archives
  Nederlandaises des Sciences Exactes et Naturelles\/} {\bf 25}, 363--552.

\bibitem[Pauli(1958)]{pauli1958theory}
{\sc Pauli, W.} 1958 {\em Theory of Relativity\/}. Dover Publications.

\bibitem[Rohrlich(2007)]{rohrlich2007classical}
{\sc Rohrlich, F.} 2007 {\em Classical Charged Particles\/}. World Scientific.

\bibitem[Spohn(2000)]{Spohn:EPL:2000}
{\sc Spohn, H.} 2000 The critical manifold of the lorentz-dirac equation. {\em
  Europhysics Letters\/} {\bf 50}~(3), 287.

\bibitem[Stahl {\em et~al.\/}(2015)Stahl, Hirvijoki, Decker, Embr\'eus \&
  F\"ul\"op]{stahl:prl}
{\sc Stahl, A., Hirvijoki, E., Decker, J., Embr\'eus, O. \& F\"ul\"op, T.} 2015
  Effective critical electric field for runaway-electron generation. {\em Phys.
  Rev. Lett.\/} {\bf 114}, 115002.

\end{thebibliography}

\end{document}